\documentclass[superscriptaddress,showkeys,showpacs]{revtex4}
\usepackage{graphicx}
\usepackage[tbtags]{amsmath}

\begin{document}
\title{ ${\cal{O}}(\alpha_s^2)$ QCD corrections to the electroproduction of hadrons with high\\ transverse momentum.}
\thanks{Partially supported by CONICET, Fundaci\'on Antorchas, UBACYT and 
ANPCyT, Argentina.}
\author{A. Daleo}
\email{daleo@physik.unizh.ch}
\affiliation{Institut f\"ur Theoretische Physik,  Universit\"at Z\"urich, \\
Winterthurerstrasse 190, CH-8057 Z\"urich, Switzerland}.
\author{D. de Florian} 
\email{deflo@df.uba.ar}
\affiliation{Departamento de F\'{\i}sica,
Universidad de Buenos Aires\\ Ciudad Universitaria, Pab.1 (1428)
Buenos Aires, Argentina}
\author{R. Sassot}
\email{sassot@df.uba.ar}
\affiliation{Departamento de F\'{\i}sica,
Universidad de Buenos Aires\\ Ciudad Universitaria, Pab.1 (1428)
Buenos Aires, Argentina}
\date{{\bf \today}}

\begin{abstract}
We compute the order $\alpha_s^2$ corrections to the one particle inclusive
electroproduction cross section of hadrons with non vanishing transverse 
momentum. We perform the full calculation analytically, 
and obtain the
expression of the factorized (finite) cross section at this order. 
We compare our results with  H1 data on forward production of 
$\pi^0$, and discuss the phenomenological implications of the rather large 
higher order contributions obtained in that case. Specifically, we analyze the 
cross section sensitivity to the factorization and renormalization scales, 
and to the input fragmentation functions, over the kinematical region 
covered by data. We conclude that the data is well described by the  
${\cal{O}}(\alpha_s^2)$ predictions within the theoretical uncertainties and without the inclusion of any physics content beyond the DGLAP approach.
\end{abstract}

\pacs{12.38.Bx, 13.85.Ni}
\keywords{Semi-Inclusive DIS; perturbative QCD.}

\maketitle

\section*{Introduction}
The precise measurement of final state hadrons in lepton nucleon deep 
inelastic scattering constitutes an excellent benchmark for different 
features of perturbative quantum chromodynamics (pQCD). These processes 
are crucially sensitive to 
the three main ingredients of pQCD: the parton content of the nucleon,
the hadronization mechanism of partons into the detected final state hadron, 
and the parton radiation before and after the interaction with the 
electromagnetic prove.

The first of these ingredients is well characterized by modern parton 
distribution functions (PDFs). The knowledge on these distributions 
have become increasingly precise as a result of two decades of high precision 
inclusive measurements, and the corresponding QCD analyses, driven by the role
of PDFs as inputs for theoretical predictions for any experiment involving 
initial state hadrons \cite{MRST04,CTEQ}. Although the high degree of accuracy 
attained by PDFs, 
less inclusive observables, sensitive to flavor combinations of PDFs other 
than those relevant in inclusive measurements, improve the insight and 
provide a further check on the universality of PDFs and on factorization. 

The second ingredient is addressed by the so called fragmentation functions 
(FF), which are rapidly evolving following the path of PDFs, but without 
attaining yet the refinement of the latter \cite{KKP,kretzer}. Most of the 
data used to determine these distributions, which come essentially from 
electron-positron 
annihilation into hadrons, give no information on how the individual quark 
flavor fragments into hadrons and leave a considerable uncertainty on the 
gluon density. For these reasons, the NLO analysis of one particle inclusive 
data is crucial for the extraction of fragmentation functions.  

The last ingredient concerns higher order QCD calculations, which have been 
explored and validated for most processes up to next to leading order (NLO) 
accuracy, and are currently being extended even beyond that point. 
For the one particle inclusive processes only very recently there has been 
progress beyond the leading order (LO) 
\cite{gluon,quark,aure,Fontannaz:2004ev,Maniatis:2004xk}. 
However, up to now there were no analytic computation of the 
${\cal{O}}(\alpha_s^2)$ corrections for the 
electroproduction of hadrons with non vanishing transverse momentum. The 
analytic computation of the ${\cal{O}}(\alpha_s^2)$ corrections allows us to 
check factorization in a direct way, which means that collinear 
singularities showing up in the partonic cross section
factorize into PDFs as required by inclusive deep inelastic scattering, 
and into FFs for electron-positron annihilation into hadrons.
As a consequence of this explicit cancellation, the resulting cross section
is finite and can be straightforwardly convoluted with PDFs and FFs in a faster
and more stable numerical code, compared to what can be usually obtained in 
numerical implementations using either the subtraction \cite{Maniatis:2004xk}
 or the slicing \cite{aure,Fontannaz:2004ev} methods.
The analytical result is still sufficiently exclusive and keeps the dependence 
on the rapidity and the transverse momentum of the produced hadron, allowing 
a detailed comparison with the experimental data.

In this paper we compute the order $\alpha_s^2$ corrections to the one 
particle inclusive leptoproduction cross section of hadrons with non 
vanishing transverse momentum. We perform the cancellation of collinear 
singularities analytically, and obtain the full expression of the 
factorized, and thus finite, cross section at this order. 
The outline of the paper is as follows: in the next section we summarize 
the relevant kinematics and details about the phase space integration for 
the O($\alpha_s^2$) contributions to the cross section, together with the 
conventions and notation adopted.
In section II we compute the corresponding real and virtual amplitudes, 
we discuss the nature of the singularities that contribute to them,  we 
analyse the factorization of collinear singularities, 
and pay special attention to the scale dependence induced in the cross 
section by this factorization procedure.
Section III deals with the phenomenological implications of the new 
corrections. 
Specifically, we compare our results with data on forward production of 
$\pi^0$, presented recently by the H1 collaboration, as an example, and 
evaluate the phenomenological implications of the rather large higher order 
corrections. Special attention is paid to the cross section sensitivity to 
the factorization scale choosen, and to the fragmentation functions input as
sources of theoretical uncertainties.  We also analyze the consequences of the 
forward pion selection on the LO and NLO underlying partonic processes, 
finding this kinematical suppression as the main reason for the 
unusual difference between the LO and NLO estimates. 
In agreement with the results obtained in \cite{aure}, we conclude that the 
data is well described by the pure Dokshitzer-Grivov-Lipatov-Altarelli-Parisi (DGLAP) ${\cal{O}}(\alpha_s^2)$ predictions 
within the theoretical uncertainties, but without need to appeal to resolved 
photon contributions, as suggested in  \cite{Fontannaz:2004ev}.

%%%%%%%%%%%%%%%%%%%%%%%%%%%%%%%%%%%%%%%%%%%%%%%%%%%%%%%%%%%%%%%%%%%%%%%%%%%%%
%%%%%%%%%%%%%%%%%%Kinematics and O(alpha_s) results%%%%%%%%%%%%%%%%%%%%%%%%%%
%%%%%%%%%%%%%%%%%%%%%%%%%%%%%%%%%%%%%%%%%%%%%%%%%%%%%%%%%%%%%%%%%%%%%%%%%%%%%

\section{Kinematics}
We begin with the kinematical characterization of the one particle inclusive 
deep inelastic scattering processes. Since the choice of variables required  
to deal with the singularity structure of electroproduction is different 
from those used in both photo-production \cite{Aurenche:1983eq,Gordon:1994wu,deFlorian:1998fq} and electroproduction 
in the very forward region \cite{gluon,quark}, in the following we discuss it in some detail. We consider the process 
\begin{equation}
l(l)+P(P)\longrightarrow l^{\prime}(l^\prime)+h(P_h)+X,
\end{equation}
where a lepton of momentum $l$ scatters off a nucleon of momentum $P$ with 
a lepton of momentum $l^{\prime}$ and a hadron $h$ of momentum $P_h$ tagged 
in the final state. Omitting target fragmentation at zero transverse momentum, 
which has been discussed at length in \cite{gluon,quark}, the cross 
section for this process can be written as 
\begin{equation}\label{eq:dsigma}
\frac{d\sigma^{h}}{dx_B\, dQ^2}=\sum_{i,j,n}\,\int_{0}^{1}d\xi\,\int_{0}^{1}d\zeta\,\int 
\mbox{dPS}^{(n)}\,\left[f_i(\xi)\,D_{h/j}(\zeta)\,\frac{
d\sigma^{(n)}_{ij}}{dx_{B}\,dQ^2\,\mbox{dPS}^{(n)}}\right]
\end{equation}
where $\sigma^{(n)}_{ij}$ is the partonic level cross section 
corresponding to the process
\begin{equation}
l(l)+i(p_i)\longrightarrow l^{\prime}(l^{\prime})+j(k_j)+
\mbox{n-1 additional partons}\,,
\end{equation}
before renormalization of the coupling constant and factorization of collinear singularities. $f_i(\xi)$ and $D_{h/j}(\zeta)$ are the bare parton densities and fragmentation functions, and dPS$^{(n)}$ the $n$-parton phase space.
$\xi$ is the proton momentum fraction carried by the parton  $i$ and 
$\zeta$ is the fraction of parton $j$ momentum taken away by the final 
state hadron. In addition to the usual DIS variables,
\begin{equation}
Q^2=-q^2=-(l^{\prime}-l)^2\,,\;\,\,\,\,\,\,x_{B}=\frac{Q^2}{2 P\cdot q}\,,\;\,\,\,\,\,\,y_e=\frac{P\cdot q}{P\cdot l}\,,\;\,\,\,\,\,\, S_H=(P+l)^2\,,
\end{equation}
we define Mandelstam variables both at parton and at hadron level:
\begin{align}
s&=(q+p_i)^2 &S&=(q+P)^2\,,\\
t&=-2\,q\cdot k_j &T&=-2\,q\cdot P_h\,,\\
u&=-2\,p_i\cdot k_j&U&=-2\,P\cdot P_h\,,
\end{align}
respectively. The above definitions imply
\begin{align}
s&=\xi\,S-Q^2\,(1-\xi)\,,&
t&=\frac{T}{\zeta}\,,& 
u&=\frac{\xi}{\zeta}\,U\,.\label{eq:ph3}
\end{align}
Notice that the $t>0$ ($T>0$) region exists only for $Q^2\neq 0$, feature 
that considerably reduces the integration region in the case of 
photo-production. The following step is the definition of suitable partonic 
variables to characterize the phase space. The choice of these variables is 
critical for the identification and further prescription of collinear 
singularities in the partonic cross section. We find particularly useful the
variables
\begin{align}
y&\equiv-\frac{u}{Q^2+s}&z\equiv\frac{(Q^2+s)(s+t+u)}
{s\,(Q^2+s+u)}\,,
\end{align}
with $y,z\in[0,1]$.
In terms of these partonic variables, the $n$-particle phase space can be
factorized as:
\begin{equation}\label{eq:dPS}
\mbox{dPS}^{(n)}=\widehat{\mbox{dPS}}^{(n-1)}\,dy\,dz\,,
\end{equation}
where $\widehat{\mbox{dPS}}^{(n-1)}$ includes the phase space of the `spectator'
partons (those that not fragment into the detected final state hadron) and 
the corresponding jacobian. 
For example for $n=3$, in $D=4+\epsilon$ dimensions we have
\begin{equation}\label{eq:dPS3}
\begin{split}
\mbox{dPS}^{(3)}&=\left(\frac{s}{4\,\pi}\right)^{\epsilon}\,
\frac{s}{(4\,\pi)^4\,\Gamma(1+\epsilon)}
\,(1-y)\,z^{\epsilon/2}\,y^{\epsilon/2}(1-y)^{\epsilon}\,(1-z)^{\epsilon/2}\,
dy\,dz\\
&\times\sin^{1+\epsilon}\beta_1\,\sin^\epsilon\beta_2\,d\beta_1\,d\beta_2\,,
\end{split}
\end{equation}
where $\beta_1$ and $\beta_2$ are the angles defined by the spectator partons
in their center of mass frame. In terms of the factorized phase space, 
eq.\eqref{eq:dsigma} reads
\begin{equation}\label{eq:s1}
\frac{d\sigma^{h}}{dx_B\,dQ^2}=\sum_{i,j,n}\,\int_{0}^{1}d\xi\,\int_{0}^{1}d\zeta\,\int_0^1 
\,dy\,\int_0^1\,dz\,\left[f_i(\xi)\,D_{h/j}(\zeta)\,\frac{
d\sigma^{(n)}_{ij}}{dx_B\,dQ^2\,dy\,dz}\right]\,,
\end{equation}
where $d\sigma^{(n)}_{ij}/dx_B\,dQ^2 dy\,dz$  is the partonic cross section 
already integrated over the spectator partons, and with the adequate  
normalization. 
Finally, changing variables from ($\xi$,$\zeta$) to hadronic transverse momentum $p_T$ and rapidity $\eta$, defined in the center of mass frame of the 
proton and the virtual photon, we find
\begin{equation}\label{eq:ss}
\frac{d\sigma^{h}}{dx_B\,dQ^2\,dp_T^2\,d\eta}=\sum_{i,j,n}\frac{e^{-\eta}\,\sqrt{S}}
{|p_T|\,(Q^2+S)}\,
\int_{e^\eta\,\frac{|p_T|}{\sqrt{S}}}
^{\frac{e^{2\eta}}{1+e^{2\eta}}}\,\frac{dy}{1-y}
\int_0^{1-\frac{y}{1-y}\,e^{-2\eta}}\,\frac{dz}{1-z}\,%\\
\left[f_i(\xi)\,D_{h/j}(\zeta)\,\frac{
d\sigma^{(n)}_{ij}}{dx_B\,dQ^2\,dy\,dz}\,\right]\,.
\end{equation}
In terms of the hadronic variables, $\xi$ and $\zeta$ are given by
\begin{align}
\xi&=\frac{Q^2\,(1-y)\,(1-z)+S\,y\,e^{-2\eta}}{(Q^2+S)\,(1-y)\,(1-z)} &
\zeta=\frac{e^{\eta}|p_T|}{\sqrt{S}\,y} \,.
\end{align}
Clearly, the transformation is singular at $y=0$, $y=1$ and $z=1$, however 
these points are excluded by  $|p_T|>0$ (notice that $|\eta|$ is bounded 
from above),
as can be seen from the limits in eq. (\ref{eq:ss}). Finally, in order 
to obtain
more compact expressions for the partonic cross sections, it turns out to be 
convenient to introduce the auxiliary variable
\begin{equation}
\varrho=\frac{x_B}{\xi}=\frac{Q^2}{Q^2+s}\,.
\end{equation}

%%%%%%%%%%%%%%%%%%%%%%%%%%%%%%%%%%%%%%%%%%%%%%%%%%%%%%%%%%%%%%%%%%%%%%%%%%%%

\section{Order $\alpha_s$ and  $\alpha_s^2$ partonic cross sections}

The partonic cross sections in eq.\eqref{eq:ss} are calculated order by order 
in perturbation theory and are related to the parton-photon squared matrix 
elements 
$\overline{H}^{(n)}_{\mu\nu}(i,j)$ for the $i+\gamma\rightarrow j+X$ processes
\begin{equation}\label{eq:psigma}
\frac{d\sigma^{(n)}_{ij}}{dx_B\, dQ^2\,dy\,dz}=\frac{\alpha_{em}^2}{e^2}
        \frac{1}{\xi\,x_B^2\,S_H^2}
        \left(Y_M (-g^{\mu\nu})+Y_L \frac{4x_B^2}{Q^2}P^{\mu}P^{\nu}
        \right)\sum_{n}\overline{H}^{(n)}_{\mu\nu}(i,j)\, .
\end{equation} 
Matrix elements are averaged over initial state polarizations, summed
over final state polarizations, and integrated over the spectator partons 
(i.e. integrated
over $\widehat{\mbox{dPS}}^{(n-1)}$). 
$\alpha_{em}$ stands for the fine structure constant and 
$e$ is the electron charge. Finally, $Y_M$ and $Y_L$ are the standard 
kinematic factors for the contributions of each photon polarization and 
are given by,
\begin{equation}
Y_M=\frac{1+(1-y_e)^2}{2\,y_e^2}\,,\:\:\:Y_L=\frac{1+4(1-y_e)+(1-y_e)^2}{2\,y_e^2}\,.
\end{equation}
The first contribution to the cross section \eqref{eq:ss} comes from the partonic 
tensor
at order $\alpha_s$, as, in the naive parton model (${\cal{O}}(\alpha_s^0)$), final 
state 
hadrons can only be produced with $|p_T|=0$ in the proton-virtual photon rest frame. 
At order $\alpha_s$, the partonic cross sections have no collinear divergences 
provided $|p_T|>0$. Up to order $\epsilon$, they are 
given by 
\begin{align}
\begin{split}\label{eq:csqq}
\frac{d\sigma^{(1)}_{qq}}{dx_B\,dQ^2\,dy\,dz}=\frac{c_q\,C_{\epsilon}}
{\xi\,x_B^2\,S_H^2}\,
C_F\,&
\Bigg\{
Y_M\,
  \left[ 
    \frac{ (\varrho+y)^2+2\,(1-\varrho-y)}{\left(1 - \varrho \right) \,
\left( 1 - y \right) }\,
    \left( 
      1 + \frac{\epsilon}{2} \,L_1
    \right) 
    +\frac{\epsilon}{2}
    \frac{{\left( \varrho - y \right) }^2}
    {\left( 1 - \varrho \right) \,\left( 1 - y \right) }
  \right]
\\
&
  +  Y_L\,
  \left[
    2\,\varrho\,y\,\left( 1 + 
      \frac{\epsilon}{2} \,L_1
    \right)
  \right]+{\cal O}(\epsilon^2)
\Bigg\}\,\delta(z)\,,
\end{split}\\
\begin{split}\label{eq:csqg}
\frac{d\sigma^{(1)}_{qg}}{dx_B\,dQ^2\,dy\,dz}=\frac{c_q\,C_{\epsilon}}
{\xi\,x_B^2\,S_H^2}\,
C_F\,&
\Bigg\{
Y_M\,
  \left[
    \frac{\left( 1 + (\varrho-y)^2\right)}{(1 -\varrho)\,y}\,
    \left( 1 + \frac{\epsilon}{2}\,L_1 \right) +
    \frac{\epsilon}{2}\,
    \frac{{\left( 1 - \varrho - y \right) }^2}
    {\left(1 - \varrho \right) \,y}
  \right]
\\
&
  +  Y_L\,
  \left[2\,
    \varrho\,\left( 1 - y \right) \,\left( 1 + 
      \frac{\epsilon }{2}\,L_1 \right)
  \right]+{\cal O}(\epsilon^2)
\Bigg\}\,\delta(z)\,,
\end{split}\\
\begin{split}\label{eq:csgq}
\frac{d\sigma^{(1)}_{gq}}{dx_B\,dQ^2\,dy\,dz}=\frac{c_q\,C_{\epsilon}}
{\xi\,x_B^2\,S_H^2}\,
T_F\,&
\Bigg\{
Y_M\,
\left[
  \frac{\left( 1 - 2\,\left( 1 - \varrho \right) \,\varrho - 
       2\,\left( 1 - y \right) \,y \right)}{\left(1 - y \right) \,y}
   \,\left( 1 + \frac{\epsilon}{2}\,L_1 \right) +
  \epsilon\,\frac{\left( \left(1 - \varrho \right) \,\varrho + 
       \left( 1 - y \right) \,y \right)}{\left( 1 - y \right) \,y}
\right]
\\
&
  +  Y_L\,
  \left[
  4\,\left( 1 - \varrho \right) \,\varrho\,
  \left( 1 + \frac{\epsilon }{2}\,L_1 \right)
  - 2\,\epsilon\,\left( 1 - \varrho \right) \,\varrho
  \right]+{\cal O}(\epsilon^2)
\Bigg\}\,\delta(z)\,,
\end{split}
\end{align} 
where
\begin{equation}
c_q=\alpha_{em}^2 2\pi\,(2+\epsilon)\,e_q^2\,,\:\:\:\:\:\:\:\:C_{\epsilon}=
        \frac{\alpha_s}{2\pi}\,f_{\Gamma}
        \left(\frac{Q^2}{4\pi\mu^2}\right)^
        {\epsilon/2},\:\:\:\:\:\:\:\:f_{\Gamma}={
        \frac{\Gamma(1+\epsilon/2)}{\Gamma(1+\epsilon)}}\,
\end{equation}
and
\begin{equation}
L_1=\log\left( \frac{(1-\varrho)(1-y) y}{\varrho}\right) \, .
\end{equation}
At order-$\alpha_s^2$, the partonic cross sections receive contributions from the 
following  
reactions:
\begin{equation}\label{eq:partreac}
\begin{array}{ll}
\mbox{Real contributions}&\left\{
\begin{array}{lcl}
\gamma+q(\bar{q})&\rightarrow& g+g+q(\bar{q})\\
\gamma+q_{i}(\bar{q_{i}})&\rightarrow&q_{i}(\bar{q_{i}})+q_{j}+\bar{q_{j}}
\,\,\,\,(i\neq j)\\
 \gamma+q_{i}(\bar{q_{i}})&\rightarrow&q_{i}(\bar{q_{i}})+q_{i}+\bar{q_{i}}\\
\gamma+g&\rightarrow& g+q+\bar{q}
\end{array}
\right.\\
\mbox{Virtual contributions}&\left\{
\begin{array}{lcl}
\gamma+q(\bar{q})&\rightarrow& g+ q(\bar{q})\\
\gamma+g&\rightarrow& q+\bar{q}
\end{array}
\right.
\end{array}
\end{equation}
where any of the outgoing partons can fragment into the final state hadron 
$h$. Order $\alpha_s$ and  $\alpha_s^2$ contributions in the very forward
region
and its singularity structure have already discussed in deep in references 
\cite{gluon,quark}. In this section we analyze the  $|p_T|>0$ region, and 
examine 
the nature of the singularities that it involves.  These contributions are 
computed in $d=4+\epsilon$ dimensions, in the Feynman gauge, and considering 
all the quarks as massless. Algebraic manipulations 
were performed with the aid of the program {\sc Mathematica} \cite{math} and 
the package {\sc Tracer} \cite{tracer} to perform the traces over the Dirac 
indices.

The order $\alpha_s^2$ partonic cross sections can be obtained from the 
corresponding
quark and gluon initiated amplitudes as in references \cite{gluon,quark}, 
taking care
of the appropriate flavor discrimination. The angular integrations can be
 performed 
with the standard techniques \cite{ert,been}, taking into account the additional
 complications  
of the one particle inclusive case: the necessity of collecting to all orders
 the potentially 
singular factors in the three particle final state integrals.
For the integrals that are known to all orders in $\epsilon$, 
this is not a problem, while for those which are only known up to a given 
order a careful treatment is required.
Once the angular integrals are performed, matrix elements are still 
distributions in three variables, $y$ and $z$ and $\varrho$, regulated by 
the parameter $\epsilon$.

At variance with the very forward case ($|p_T|=0$), where the integration 
over final states leads to overlapping singularities along various curves in 
the residual phase space, here the only remaining singularities are found at 
$z=0$ and thus they can be dealt with the standard method. 
After combining real and virtual contributions to a given partonic process, 
the cross section can be written as 
\begin{eqnarray}\label{eq:psigma2}
\frac{d\sigma^{(2)}_{ij}}{dx_B\,dQ^2\,dy\,dz}&=& \frac{c_q\,C_{\epsilon}^2}{\xi\,x_B^2\,S_H^2}\,
\left\{\frac{1}{\epsilon}\,{\cal{P}}^{(2)}_{1\, ij}(\varrho,y,z)\,
        +C^{(2)}_{ij}(\varrho,y,z)+{\cal O}(\epsilon)
        \right\}
\,, 
\end{eqnarray}
where the coefficient of the single poles, 
${\cal{P}}^{(2)}_{1\,ij}(\varrho,y,z)$, as well
as the finite contributions $C^{(2)}_{ij}(\varrho,y,z)$, include `delta' and `plus' 
distributions in $z$.
The IR double poles present in the individual real and virtual contributions cancel 
out 
in the sum, providing the first straightforward check on the angular integration 
of real amplitudes and
the loop integrals in the virtual case. In the real terms, the above 
mentioned double poles come from the product of a pole arising in the 
integration over the spectators phase space (i.e. integration
over $\beta_1$ and $\beta_2$ in \eqref{eq:dPS3}) and a single pole coming from the 
expansion of
$z^{-1+\epsilon}$ factors. Double poles in the virtual contributions always arise 
from loop 
integrals.

The remaining singularities, contributing to the single pole, are of UV and 
collinear origin. The 
former are removed by means of coupling constant renormalization, whereas the latter 
have to be factorized in the redefinition of 
parton densities and fragmentation functions. The redefinition
of parton densities is exactly the same as in totally inclusive DIS whereas
fragmentation functions are renormalized as they are in one-particle inclusive
electron-positron annihilation. Typical expressions for renormalized parton 
densities and fragmentation functions, up to order $\alpha_s^2$ and 
in the $\overline{MS}$ factorization scheme, can be found, for example, in
Refs.\cite{zijli} and \cite{rijken} respectively. Factorization of 
collinear singularities and cancellation of the UV ones, then impose
\begin{equation}\label{eq:facto}
{\cal{P}}^{(2)}_{1\,ij}(\varrho,y,z)=2\,(C_{lj}^{(1)}\otimes P^{(0)}_{li} 
+C_{ik}^{(1)}\otimes P^{(0)}_{jk})-\beta_0\,C_{ij}^{(1)}
\end{equation}
where the $C_{ik}^{(1)}$ correspond to the finite (${\cal O}(\epsilon^0)$) terms 
in the ${\cal{O}}(\alpha_s)$ partonic cross sections of eqs. 
\eqref{eq:csqq}, \eqref{eq:csqg} and \eqref{eq:csgq}. 
$P_{ij}$ are the standard  LO Altarelli-Parisi kernels, and
$\otimes$ denotes the appropriate convolution coming from the factorization
recipes. 

The factorized, and thus finite, partonic cross sections have
terms proportional to $\delta(z)$, terms containing `plus' distributions,
and purely functional contributions. The logarithmic `plus' contributions have 
their origin in the multiple emission of soft-gluons and can therefore be predicted 
by taking the order $\alpha_s$ expansion of the corresponding resumed cross-section. 
For a partonic subprocess initiated by a parton $i$, 
where a parton $j$ fragments, and with a gluon and a parton $s$ as spectators, 
$\gamma+i\rightarrow g+j+s$, the result is
\begin{equation}
\left.\frac{d\hat{\sigma}^{(2)}_{ij}}{dx_B\,dQ^2\,dy\,dz}\right|_+=
\frac{d\hat{\sigma}^{(1)}_{ij}}{dx_B\,dQ^2\,dy\,dz} 
\left[1+\frac{\alpha_s}{2\pi}\left( \frac{\mbox{ln}z}{z}\right)_+ 
\left(4\,C_i+4\, C_j-2\,C_s \right)\right]
\end{equation}
where the general color factor $C_k$ corresponds to $C_F$ if $k$ is a quark and to 
$C_A$ if it 
is a gluon. The agreement with this prediction provides a further test on our 
results. 

Since the factorized coefficients get contributions from both the real and 
virtual process at order $\alpha_s^2$, together with finite terms coming from
the renormalization and factorization procedure, their explicit expressions
are considerably long and thus are omitted here \footnote{The explicit 
expressions for the factorized 
coefficients can be obtained from the authors upon request.}.

Notice that the renormalization and factorization processes introduce scale 
dependent terms in the final cross section which partially cancel the scale
dependence induced by the coupling constant, parton densities, and 
fragmentation functions. The structure of 
these terms follows that of the factorization contributions in 
eq.\eqref{eq:facto}. 
 \begin{equation}
-2\left[C_{lj}^{(1)}\otimes P_{li}\, \,\mbox{ln}\left(\frac{M_F^2}{Q^2} \right)+
C_{ik}^{(1)}\otimes P_{jk}\, \, \mbox{ln}\left(\frac{M_D^2}{Q^2} \right)\right]+
\beta_0 C_{ij}^{(1)}\,\, \mbox{ln}\left(\frac{M_R^2}{Q^2} \right)
\end{equation}
where  $M_R$, is the renormalization scale and  $M_F$ and $M_D$ are factorization 
scale for parton densities and fragmentation functions, respectively. 

In order to visualize the magnitude of the higher order corrections, in 
Figure 1 we show the ratio between the order $\alpha_s^2$ and $\alpha_s$ 
cross sections for $\pi^0$ production (K-factor) as a function of $p_T$ for 
$Q^2=200\, \mbox{GeV}^2$, integrated over rapidity and for different 
values of $x_B$ in the kinematically
allowed range. As input parton densities and 
fragmentation functions we choose the MRST02 \cite{MRST02} and KKP \cite{KKP},
NLO and LO sets, respectively. In the following we refer to the convolution 
of ${\cal{O}}(\alpha_s^2)$ cross sections and NLO densities as NLO 
prediction, whereas ${\cal{O}}(\alpha_s)$ cross sections convoluted with LO 
densities define 
the LO estimate.
\setlength{\unitlength}{1.mm}
\begin{figure}[hbt]
\includegraphics[width=9cm]{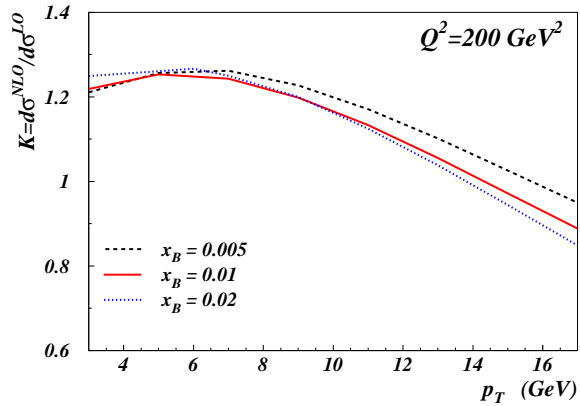}
\caption{K-factor as a function of $p_T$.}\label{fig:k-factor}
\end{figure} 
The renormalization and factorization scales $M_R^2$, $M_F^2$ and $M_D^2$ 
were taken to be the average between the two main physical scales of the 
process, namely the transverse momentum of the final state particle and the 
virtuality of the photon as
\begin{equation}
M_R^2=M_F^2=M_D^2=\frac{Q^2+p_T^2}{2}\, .
\end{equation}
The K-factor exhibits the characteristic behavior of higher order 
corrections; they increase at low transverse momentum and also at low $x_B$. 
At very high $p_T$, where the LO estimate becomes larger than the NLO prediction, threshold effects become dominant and the perturbative expansion at fixed 
order in the coupling constant  is not expected to be reliable.

\begin{figure}[hbt]
\includegraphics[width=9cm]{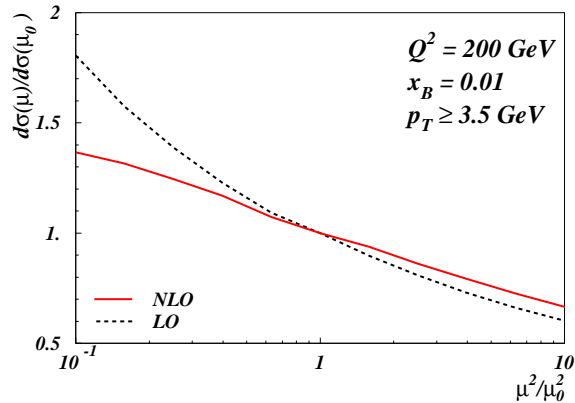}
\caption{Scale dependence of the electroproduction cross section.}\label{fig:mumu}
\end{figure}

The dependence on the particular choice for the factorization and 
renormalization scales is expected to be weaker at NLO than at LO. 
In Figure  \ref{fig:mumu} we show this dependence plotting the rate 
between the cross section evaluated at an arbitrary scale $\mu^2$ and the 
cross section at $\mu^2_0=(Q^2+p_T^2)/2$ as function of the rate 
$\mu^2/\mu^2_0$.
\setlength{\unitlength}{1.mm}
As in the previous plot $Q^2=200\, \mbox{GeV}^2$,  but $x_B=0.01$ and we 
integrate over the allowed $p_T$ range, starting from 
$|p_T|>3.5\,\mbox{GeV}$. As expected, the scale dependence is milder for 
the NLO estimate, although it is not negligible.

Notice that our NLO estimate focus on the `direct' coupling of photons 
to partons, without taking into account the  `resolved' photon contributions,
as those computed with virtual photon parton densities. These contributions 
have been carefully analyzed in references \cite{aure,Fontannaz:2004ev}.

\section{Phenomenology}

Recently the H1 \cite{Aktas:2004rb} collaboration has presented an improved 
measurement of the production of neutral pions in collisions between 
$27.6\mbox{ GeV}$  positrons and $820\mbox{ GeV}$ protons. Neutral pions are 
required to be produced within a small angle $\theta_{\pi}$ from the proton 
beam in the laboratory 
frame ($\theta_{\pi} \in[5^o,25^o]$), with an energy fraction 
$x_{\pi}=E_{\pi}/E_P>0.01$ and $2.5<p_T<\,15\mbox{ GeV}$. The data confirmed 
previous measurements which suggested that QCD LO predictions underestimate 
the cross section at low $x_B$ \cite{H1viejo}. On the other side, 
predictions based on BFKL dynamics \cite{BFKL}, or on a large 
virtual photon content \cite{Jung} seemed to provide better descriptions.

The disagreement between the H1 data and estimates based on  
${\cal{O}}(\alpha_s)$ 
cross sections convoluted with LO parton densities and fragmentation 
functions can be as large as an order of magnitude, depending on the 
particular kinematical region. This discrepancy is far larger than the 
typical K-factor found in the previous section, what suggests the onset of 
a physical mechanism different to leading or next to leading order 
DGLAP dynamics. 

However, several non-negligible effects are present at the particular 
kinematical regime of the experiment, which are responsible for a large 
difference between the LO and NLO estimates. The first one is the
stringent cut on the pion production angle in H1 data, which suppresses LO 
and NLO contributions in a  different way. The  
suppression of LO configurations is proportionally bigger than for NLO,
 implying an 
effective K-factor much  larger than the one found for the cross 
section without cuts.
The second important feature is the rather low value of the scales involved 
($p_T$ and $Q^2$) which enhance the uncertainty due to the 
particular choice for the factorization scale, even in the NLO calculation, 
as it has been pointed in \cite{aure}. 
This is particularly significant for the lowest $Q^2$ bins.
Finally, there is also a large uncertainty 
factor in the theoretical prediction coming from fragmentation functions. 
Although fragmentation functions reproduce fairly well $e^+e^-$ annihilation 
into hadrons, they show large differences when they are used to compute deep inelastic semi-inclusive cross sections.

\begin{figure}[h]
\includegraphics[width=8cm]{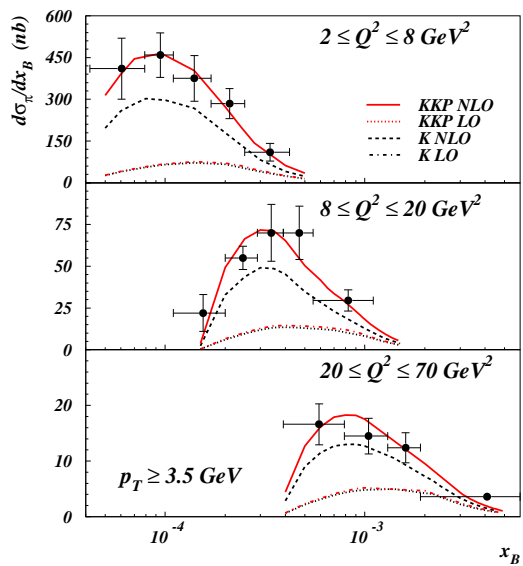}
\caption{LO and NLO cross sections, including experimental cuts as explained in the text,
as a function of $x_B$. H1 data \cite{Aktas:2004rb} for the range $p_T\ge 3.5\mbox{ GeV}$
are also shown.
}\label{fig:x_Bj}
\end{figure}

\begin{figure}[h]
\includegraphics[width=8cm]{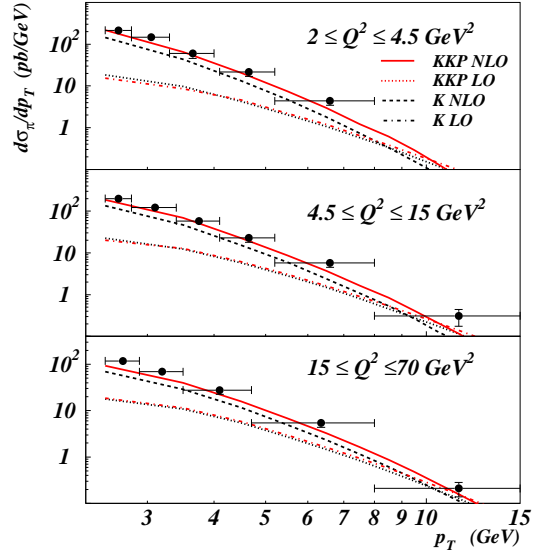}
\caption{Cross section as a function of $p_T$, data and cuts as in Figure \ref{fig:x_Bj}.
}\label{fig:pT}
\end{figure}

In Figures \ref{fig:x_Bj} and \ref{fig:pT} we show the LO and NLO 
predictions for the electroproduction of neutral pions as a function of 
$x_B$ and $p_T$, respectively, in the kinematical range of the H1 
experiment, together with the most recent data for the range 
$p_T\ge 3.5\mbox{ GeV}$. 
The cross sections are computed as described in the previous sections, 
applying H1 cuts and using MRST02 parton densities \cite{MRST02}.
Similar results are found using other sets of modern PDFs. For the 
input fragmentation functions, we use two different sets, the ones from 
reference \cite{KKP} denoted as KKP and those from \cite{kretzer} referenced 
as K. We set the renormalization and factorization scales as in eq. (27) and 
we compute $\alpha_s$ at NLO(LO)  fixing $\Lambda_{QCD}$ as in the MRST 
analysis, such that $\alpha_s(M_Z)=0.1197(0.130)$.  

\setlength{\unitlength}{1.mm}

\setlength{\unitlength}{1.mm}

The plots clearly show some of the features mentioned above. 
On the one hand, the NLO cross sections are much larger than the LO ones,
even by the required order of magnitude in certain kinematical regions, 
 once the forward $\pi^0$ selection applied by H1 is implemented. The 
position of the maximum for the $x_B$ distribution is also shifted to 
lower $x_B$ values, agreeing nicely with the experimental shape.
Cross sections differential in $p_T$ show similar features, however the 
difference between LO and NLO decreases as $p_T$ increases.
 
The uncertainty due to the choice of a fragmentation functions set is also 
quite noticeable, this fact driven by the different gluon content of the 
two sets considered here. Low $Q^2$ bins seem to prefer KKP set, which 
have a larger gluon-fragmentation content, whereas for larger $Q^2$ both sets 
agree with the data within errors. LO estimates show a much smaller 
sensitivity 
on the choice of fragmentation functions, since gluon fragmentation does not 
contribute significantly to the cross section at this order.

\setlength{\unitlength}{1.mm}
\begin{figure}[hbt]
\includegraphics[width=9cm]{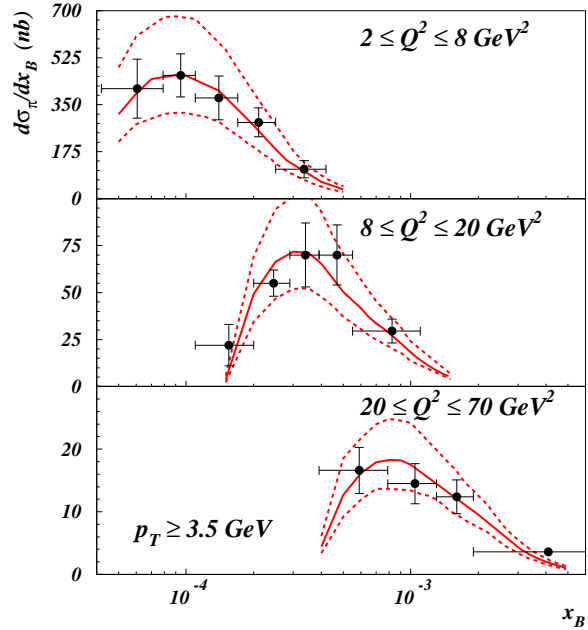}
\caption{NLO cross sections as a function of $x_B$. The central (solid) line
corresponds to setting the factorization and renormalization scales to 
$(Q^2+p_T^2)/2$ and the upper and lower (dashed) ones to $(Q^2+p_T^2)/4$ and 
$Q^2+p_T^2$ respectively.  
}
\label{fig:scale3}
\end{figure}
As mentioned, the dependence of the cross section in the choice for the 
renormalization and factorization scale is also an important source of 
uncertainty even at NLO.
In Figure \ref{fig:scale3} we show the NLO prediction with the standard choice 
for the scale, MRST02 parton densities and KKP fragmentation functions 
(solid line) as in Figure \ref{fig:x_Bj}, together with H1 data and the 
estimates with a scale twice as large (lower dashes) and another scale 
half of the former (upper dashes). 

\begin{figure}[hbt]
%\begin{picture}(30,50)(0,0)
%\put(-30,-35){\mbox{\epsfxsize8.cm\epsffile{FIG5.ps}}}
%\end{picture}
\includegraphics[width=9cm]{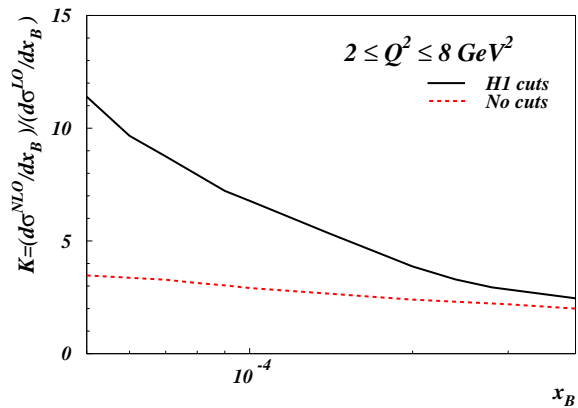}
\caption{K-factors for the lowest $Q^2$ bin of Figure \ref{fig:x_Bj} 
as a function of $x_B$ with and without the experimental cuts.}
\label{fig:kfacx}
\end{figure}

\begin{figure}[b!]
%\begin{picture}(30,75)(0,0)
%\put(-30,-15){\mbox{\epsfxsize8.cm\epsffile{FIG6.ps}}}
%\end{picture}
\includegraphics[width=9cm]{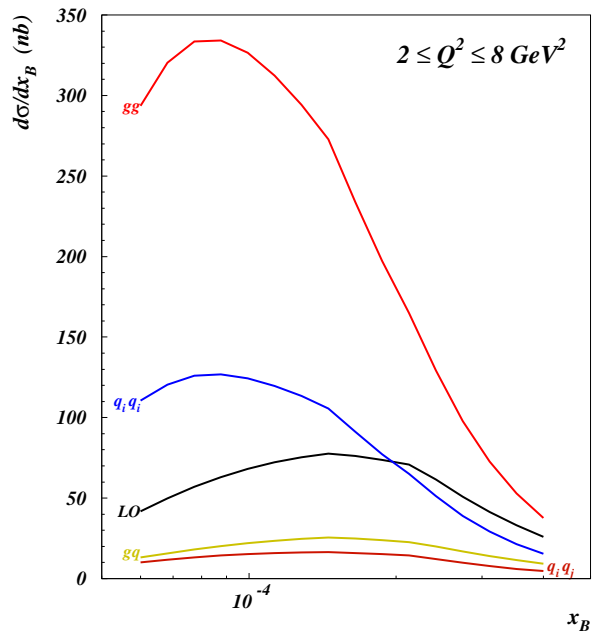}
\caption{Contributions to the cross section discriminated by 
the underlying partonic process for the lowest $Q^2$ bin of Figure \ref{fig:x_Bj}, 
including experimental cuts. 
Processes $qg$ and $q\bar{q}$ give negligible contributions and are not shown
in the plot.}
\label{fig:discri}
\end{figure}

Finally, in order to illustrate the effects of the forward selection criteria, 
in Figure  \ref{fig:kfacx} we show the effective K-factor
for the lowest $Q^2$ bin, with and without taking into account the 
constraints on $\theta_{\pi}$ and $x_{\pi}$. 
\setlength{\unitlength}{1.mm}
Although the low values of $x_B$ and $Q^2$ lead to a very large K-factor, 
the forward selection, typically enhances it by a factor of three.
Notice that  the process $\gamma+g\rightarrow g+q+\bar{q}$ becomes active 
at  ${\cal{O}}(\alpha_s^2)$ and indeed turns out to be 
responsible for most of the correction, as it is illustrated in Figure
\ref{fig:discri}, where we show the different contributions to the cross 
section discriminated by the underlying partonic process.
\setlength{\unitlength}{1.mm}

The rather large size of the K-factor can, then, be understood  
as a consequence of the opening of a new dominant (`leading-order') channel,
 and not to the `genuine' increase in the partonic cross 
section that might otherwise threaten perturbative stability. The dominance of
the new channel is due to the size of the gluon distribution 
at small $x_B$ and to the fact that the H1 selection cuts highlight
the kinematical  region dominated by the $\gamma+g\rightarrow g+q+\bar{q}$ 
partonic process. In particular, without the experimental cuts for the final 
state hadrons, the $gg$ component represents less than $25\%$ of the total 
NLO contribution at small $x_B$, which is dominated by the $gq$ subprocess.
%
%In Figure \ref{fig:discri}  we show the different contributions to the cross 
%section discriminated by the underlying partonic process. Notice that at very 
%small $x_B$ the $gg$ term can be by itself several times larger than the LO 
%contribution, remaining larger or comparable even for higher $x_B$ values. 
%
The forward selection is also responsible of the scale sensitivity of the
cross section, as it supresses large components  with small scale
dependence whereas it stresses components as $gg$ whose scale dependence would
be partly canceled only at NNLO.

\section{Conclusions}

We have presented the analytical calculation of the differential cross section
for semi-inclusive production of a hadron, with non vanishing transverse 
momentum, in DIS at next-to-leading-order in QCD.  
As for any semi-inclusive process, the necessity of integrating the phase 
space of the unobserved particles, but keeping the full dependence on the 
variables characterizing the final state hadron (and thus of the parton from 
where it comes), makes the computation of higher order corrections much more 
involved than the inclusive case. In the present case we showed that, with a 
suitable parameterization of the phase space, the necessary integrations can be 
performed analytically and the remaining singularities can be dealt with 
standard prescription recipes, without the need of substraction or phase space 
slicing methods.

We found that the order $\alpha_s^2$ corrections are important, leading to 
large K-factors. The main contributions to these corrections come from 
the partonic subprocess $ \gamma + g \rightarrow g+q+\bar{q}$ which appears
for the first time at that order. The appearance of new channels
also leads to quite a significant factorization scale dependence even at the 
NLO level. 

Concerning the phenomenological consequences of our results, we compared them 
with recent data coming from the H1 experiment at HERA \cite{Aktas:2004rb}. 
Within the uncertainties arising from the scale dependence and the particular 
sensitivity of the results to the gluon hadronization mechanism, parameterized 
in the fragmentation functions, we found a very good agreement between data 
and theoretical expectations for both the $x_B$ and $p_T$ distributions. 
In particular, fixing the factorization and
renormalization scales as just the average between the photon virtuality and 
the transverse momentum of the final state hadron, both distributions are 
well described by purely DGLAP evolution. We also found that the experimental
cuts applied to the H1 data play a crucial role, boosting the NLO corrections,
and thus explaining the unusual poor description of the LO estimate. Finally, 
our results are in agreement with those obtained previously by numerical 
methods \cite{aure,Fontannaz:2004ev}.

The large K-factors and the significant factorization scale dependence, both
related to the opening of new channels at NLO, suggest the
presence of non negligible NNLO effects. This feature, altogether with the 
fact that data is reasonably described by NLO estimates within uncertainties, 
obliterate any further effect that might contribute, like the resolved 
component of the photon or those coming from BFKL dynamics in the present 
experiments.

\section{Acknowledgements}

 We warmly acknowledge C. A. Garc\'{\i}a Canal  for comments and suggestions. 
The work of one of us (A.D.) was supported in part by the Swiss National
Science Foundation (SNF) through grant No. 200021-101874.

\end{document}